\documentstyle[amssymb,prb,aps]{revtex}
\begin{document}
\twocolumn[\hsize\textwidth\columnwidth\hsize\csname
@twocolumnfalse\endcsname

\title{Charge disproportionation in YNiO$_{3}$ : ESR and susceptibility study}
\author{M. T. Causa\thanks{Corresponding author;
email: causa@cab.cnea.gov.ar}, R. D. S\'{a}nchez, and M. Tovar}
\address{Centro At\'{o}mico Bariloche, CNEA, Av. Ezequiel Bustillo 9500, (8400) \\
San Carlos de Bariloche, R\'{\i}o Negro, Argentina.}
\author{J. A. Alonso and M. J. Mart\'{\i}nez-Lope.}
\address{Instituto de Ciencia de Materiales de Madrid, CSIC, Cantoblanco, E-28049
\\Madrid, Spain. } \maketitle

\begin{abstract}
We present a study of the magnetic properties of YNiO$_{3}$ in the
paramagnetic range, above and below the metal-insulator (MI)
transition. The dc susceptibility, $\chi _{dc}$ (measured up to
1000 K) is a decreasing function of T for $T >$150 K (the N\'{e}el
temperature) and we observe two different Curie-Weiss regimes
corresponding to the metallic and insulator phases. In the
metallic
phase, this behaviour seems to be associated with the small ionic radius of Y%
$^{3+}$. The value of the Curie constant for T$<$ T$_{MI}$ allows
us to discard the possibility of Ni$^{3+}$ localization. An
electron spin resonance (ESR) spectrum is visible in the insulator
phase and only a fraction of the Ni ions contributes to this
resonance. We explain the ESR and $\chi _{dc}$ behaviour for T $<$
T$_{MI}$ in terms of charge disproportionation of the type 2Ni$%
^{3+}\rightarrow $ Ni$^{2+}$+Ni$^{4+},$ that is compatible with the
previously observed structural transition across T$_{MI}$.

\end{abstract}

\vskip2pc] \narrowtext

\section{Introduction}

The perovskite system RNiO$_{3}$ (with R trivalent rare-earth or
Y) is a very attractive family of compounds around which an
intense research activity was developed in the 90\'{}s. The most
studied property in these oxides was the metal-insulator (MI)
transition\cite{Torrance} from the high temperature metallic
material to a charge transfer insulator, at a temperature
T$_{MI}$. As is seen in the phase diagram of Fig. 1, this
temperature increases when the ionic radius of R$^{3+}$ (r$_{R}$)
diminishes. In the insulator phase the oxides show
antiferromagnetic (AFM) order, with a complex structure, below a
N\'{e}el temperature (T$_{N}$) also dependent on the r$_{R}$ size. The
AFM structure was firstly explained in terms of localization of
the Ni$^{3+}$-$eg$ electrons and orbital ordering, Jahn-Teller
driven.\cite{Juan} In spite of all the performed studies, the
exact nature of the metal-insulator transition is not completely
understood\cite{Alonso1} and a model including structural,
magnetic and transport properties for all the RNiO$_{3}$ series is
still lacking. The difficulties found in the synthesis of
RNiO$_{3}$, partially related to the stabilization of the
Ni$^{3+}$, are more severe for the smaller r$_{R}$ compounds\cite
{Alonso1}$^{,}$\cite{Demazeau} and only recently,\cite{Alonso2}
with new synthesis techniques, the samples with R = Dy, Y, Ho, Er,
and Lu could be studied. Besides, improvements in the
characterization of the structural changes\cite{Alonso1}
accompanying the MI transition performed by synchrotron x-ray
diffraction (SXRD) techniques allowed a more precise description
of the different crystalline phases.

In this paper we present a study of the YNiO$_{3}$ oxide. Due to the non
magnetic character of Y, the magnetic properties of the Ni lattice can be
studied in YNiO$_{3}$ without interference with other magnetic species.
Notice that the well known LaNiO$_{3}$, where R = La is also non magnetic,
was found to be metallic for all T $\geq $ 1.4 K, without signatures of any
magnetic ordering.\cite{Rodo1}

In a previous work, Alonso et al.\cite{Alonso1} determined the
metal-insulating transition for YNiO$_{3}$, using differential scanning
calorimetry, obtaining T$_{MI}$ = 582K. They observed a structural
transition from the orthorhombic $Pbnm$ phase (above T$_{MI}$) to a
monoclinic $P21/n$ phase (below T$_{MI}$). This transition has been assumed
to be driven by Ni charge disproportionation in the insulating phase \cite
{Alonso1}$^{,}$\cite{Alonso2} where 2Ni$^{3+}\rightarrow $ Ni$^{3+\delta }$
+ Ni$^{3-\delta }$ ($\delta \approx \delta $\'{}$\approx 0.3$) . In this
interpretation, the two Ni species coexist in the insulating phase,
localized in two different crystalline sites characterized by expanded (Ni1
site) and contracted (Ni2 site), slightly distorted, NiO$_{6}$ octahedra.
The AFM spin configuration at low temperature was also determined by neutron
diffraction in the magnetically oredered phase,\cite{Alonso1} finding
different magnetic moments that were associated with the two Ni sites. In
this work we study the electron spin resonance (ESR) properties of YNiO$_{3}$%
. The aim of our work is to describe the changes in the magnetic properties
associated to the MI transition. In the metallic phase all the Ni ions are
expected to be in the low-spin (LS) Ni$^{3+}$ configuration ($%
3d^{7},t_{2g}^{6}e_{g}^{1}$). The absence of an ESR spectrum that we found
in the metallic phase is compatible with this Ni$^{3+}$ configuration where
the $e_{g}$ electrons are itinerant and the localized $t_{2g}$ electrons
have compensated spins. The same behavior (lack of ESR) has been observed%
\cite{Rodo-Tesis} for the metallic LaNiO$_{3}$ in the whole temperature
range. The ESR spectrum present in the paramagnetic (PM) insulating phase of
YNiO$_{3}$ is, as we discuss in this paper, a consequence of the Ni charge
disproportionation\cite{Alonso1} and constitutes a direct experimental
evidence of this effect. $\ $

\section{ Experimental details}

Powder YNiO$_{3}$ samples were prepared under high-pressure and
high-temperature conditions. Stoichiometric mixtures of
Y$_{2}$O$_{3}$ and Ni(OH)$_{2}$ powders were ground with 30\% of
KClO$_{4}$. The mixture was packed into a 8 mm-diameter gold
capsule, placed in a cylindrical graphite heater. The reaction was
carried out in a piston-cylinder press, under a pressure of 20
kbar at 1173 K for 20 min. The reaction product, in the form of
blackish dense polycrystalline pellets, was ground and the
resulting powder washed with water to dissolve KCl. The sample was
characterized by laboratory X-ray diffraction, with Cu K$_{\alpha
}$ radiation. The temperature dependence of the dc-susceptibility,
$\chi _{dc}$(T), was measured with a SQUID magnetometer for T $<$
300K and with a Faraday balance magnetometer in the range 250 K -
1000 K. The ESR spectrum as a function of T was measured with a
Bruker spectrometer operating at a frequency $\nu \approx $ 9.3
GHz. We detected the derivative spectrum and obtained the T
dependence of the three ESR parameters: resonance field
H$_{0}$(T), peak-to-peak linewidth $\Delta $H$_{pp}$(T), and the
double integrated intensity I$_{esr}$(T). Notice that I$_{esr}$
can be approximated by ($\Delta $ Hpp)$^{2}\times $ h$_{pp}$,
where h$_{pp}$ is the distance between peaks of the derivative
spectrum. This method was preferred to the double integration for
low signal-to-noise ratio spectra, in order to minimize possible
spurious effects due to the cavity background. A standard Bruker
X-band cavity was used for the T$<$ 400 K measurements while a
4114HT cavity was employed in order to reach the high temperature
range (up to 800 K in this case). The spectrum intensity was
compared to that of well characterized standard samples (MnF$_{2}$
single crystal and Gd$_{2}$BaCuO$_{5}$ powder\cite{Goya}) in order
to determine the ESR susceptibility, $\chi _{esr}\propto $
I$_{esr}$, that should be equal to $\chi _{dc}$ if all the
magnetic species contribute to the resonance. The detected lines
were of relatively low intensity and 100 mg samples were used in
order to obtain low noise spectra.

\section{ Results and discussion}

In Fig. 2 we show $\chi _{dc}$(T). Our results are in agreement with the
pioneering experiment by Demazeau et al.\cite{Demazeau} in the range of
their measurements (70 K / 480 K). At T = 150 K, $\chi _{dc}$(T) shows a
peak attributed\cite{Demazeau} to an AFM transition. As seen in Fig.2, at
temperatures well below T$_{N}$, a Curie-like contribution to $\chi _{dc}$
is observed, as in other perovskites,\cite{Co1}$^{,}$\cite{La1} whose nature
is unknown and could be originated in minority phases ( 0.5\%). Measurement
at 20 K of the magnetization as a function of H (up to 50 kG) showed a
linear dependence, as is usual in AFM materials, although with a very small
uncompensated magnetization, M$_{0}$ 10$^{-4}$ $\mu _{B}$/fu.

In the inset of Fig.1 we show $\chi _{dc}^{-1}$(T) vs. T. In the
paramagnetic (PM) range (T $\geq $ T$_{N}$) $\chi _{dc}$(T)
follows a Curie-Weiss (CW) law. For T$>$T$_{MI}$, a Curie constant
C = 0.79(2) emu-K/mol was obtained. Around T$%
_{MI} $ a change of slope was observed and C = 0.90(2) emu-K/mol
corresponds to T$<$T$_{MI.}$ These values of C give, respectively,
effective magnetic moments $\mu _{eff}$ = 2.5 $\mu _{B}$ and 2.7
$\mu _{B}$. The simplest picture for the MI transition would be
that of LS Ni$^{3+}$ ions with itinerant $e_{g}$ electrons in the
metallic phase becoming fully localized below T$_{MI}$. In
metallic oxides we may describe $\chi _{dc}$(T) in terms of a sum
of a T-independent ( $\chi _{0}$) and a CW-like ( $\chi _{CW}$)
susceptibilities\cite{Tovar2} corresponding to the itinerant and
localized electrons, respectively. For RNiO$_{3}$ these should be
the $e_{g}$ and $t_{2g}$ electrons of Ni$^{3+}$. In this case only
a temperature independent susceptibility $\chi _{0}$ is expected
for T$>$T$_{MI}$ since the localized $t_{2g}$ electrons have fully
compensated spins. This behavior was indeed observed\cite{Rodo1}
in LaNiO$_{3}$. Instead, as was mentioned above, a CW
susceptibility is measured in the
metallic phase of YNiO$_{3}$. A similar situation occurs\cite{Perez}$^{,}$%
\cite{Rodo2} in SmNiO$_{3}$ and in La$_{1-x}$Eu$_{x}$NiO$_{3}$ (x $\geq $
0.6) after the subtraction of the Sm$^{3+}$ and Eu$^{3+}$ contribution\cite
{Van Vleck} to the susceptibility. It is interesting to notice that in the La%
$_{1-x}$Eu$_{x}$NiO$_{3}$ series,\cite{Rodo2} a transition from a
independent to a CW behavior is observed as a function of x. Up to
x $\approx $ 0.4, $\chi _{dc}$ = $\chi _{0}$ and, for larger
values of x, a CW component becomes increasingly important. These
Eu concentrations correspond to the region of the phase diagram
(see Fig. 1) characterized by the existence of an extended
PM-insulator phase (T$_{MI}>$ T$_{N}$). Then, it seems that the
magnetic behavior in the metallic phase of RNiO$_{3}$ compounds
is, as other magnetic and transport properties, strongly dependent
on the size of the ionic radius r$_{R}$. This is shown in Fig. 1.

Below T$_{MI}$, the experimental results also fail to validate the simple
model that suggest that in the PM insulating phase LS Ni$^{3+}$ are fully
localized. In this case a CW law with C = 0.4 emu-K/mol (S = $\frac12$, $\mu
_{eff}$ = 1.7 $\mu _{B}$) is expected, value much smaller than the measured
one. We will discuss this issue after the analysis of the ESR results.

In Fig. 3 we show the ESR spectrum taken at different
temperatures. Between 150 K and 600 K, the spectrum consist of a
single lorentzian line with a constant g-factor, g = 2.16(2). At T
$\approx $ 150 K this line suddenly broadens up, shifts to lower
fields and is no longer visible for T $<$ 145 K. This behavior is
generally observed in powder samples below the AFM transition.
This is due to the splitting, in several anisotropic branches, of
the PM single-valued function $\nu $ = (g $\mu _{B}$/h)H and to
the opening of a T dependent energy-gap $\nu $(H=0) =
E$_{g}$(T)/h. These effects combine in polycrystalline samples to
give progressively broader and, consequently, lower amplitude
spectra that are finally lost within the noise. Besides, when T
diminishes, E$_{g}$(T) increases and for temperatures where
E$_{g}>$ \mbox{$>$} h$\nu $ the spectrum is not detected even in
single crystals. Our observation corroborates that the resonant Ni
ions order AFM below T$_{N}$ = 150 K in coincidence with the dc
peak temperature. On the other hand, for temperatures approaching
the MI transition, the line intensity also diminishes (see Fig. 3)
but in this case both the linewidth and the resonance field remain
approximately constant. Above 600 K the line extinguishes
completely.

In Fig. 4 we plot the measured T dependence of $\Delta $H$_{pp}$.
For the range T$_{N}<$ T $<$450 K the linewidth increases slowly
showing a linear dependence $\Delta $H$_{pp}$(T) = a + bT with a
$\approx $ 1300 G and b $\approx $ 1.5 G/K. Above 450 K the
linewidth stops to increase, a narrowing process installs, and
$\Delta $H$_{pp}$ is reduced by 50\% at 560 K. In Fig. 2 we show
$\chi _{esr}$(T) that is decreasing with T in all the range
150/600 K: between 150K and 450K we observe a behavior similar to
that of $\chi _{dc}$(T), for T $\approx $ 450 K a change in the
slope occurs, and above this temperature $\chi _{esr}$ goes
rapidly to zero.

The ESR behavior can be explained taking into account the structural changes
of the material. Above T$_{MI}$, in the orthorhombic phase, all the Ni ions
are Ni$^{3+}$ and no resonance is visible. In the monoclinic phase, instead,
it is very interesting to notice that, below T$_{MI}$, gradually more and
more Ni ions contribute to the spectrum. In Fig. 2 we compare $\chi _{esr}$
with $\chi _{dc}$, finding that $\chi _{esr}$ is, for all temperatures,
lower than $\chi _{dc}$ being $\chi _{esr}\approx $ 0.5 $\chi _{dc}$ at T $%
\approx $ 150K. This indicates that only part of the Ni ions contribute to
the ESR. The g-factor value\cite{Nota} and the linear dependence\cite{Causa1}
found for $\Delta $H$_{pp}$(T) suggest that the resonant species may be Ni$%
^{2+}$. These facts point to a charge disproportionation: 2Ni$%
^{3+}\rightarrow $ Ni$^{2+}$+Ni$^{4+}$ at the MI transition. If this were
the case, the contracted octahedra would be occupied by low-spin Ni$^{4+}$ (3%
$d^{6}$ $t_{2g}^{6}$) ions, having S = 0 ground state that would contribute
to $\chi _{dc}$ through their excited levels, as in the case of the
isoelectronic Co$^{3+}$. It is known\cite{Co1} that the perovskite LaCoO$%
_{3} $ displays a T dependent magnetic susceptibility, in spite of the non
magnetic character of the Co$^{3+}$ ground state. This magnetic behavior has
been modeled\cite{Co1} including contributions from the Co$^{3+}$ high-spin
and intermediate-spin excited levels.\cite{Bleaney} On the other hand,
attempts to observe the LaCoO$_{3}$ ESR spectrum were unsuccessful.\cite
{Rodo3} Therefore, our results suggest that (as in the case of Co$^{3+}$)
excited levels of Ni$^{4+}$ contribute to $\chi _{dc}$ but are invisible to
the ESR experiment. Supporting the strong interdependence of ESR,
dc-susceptibility, and structural properties in YNiO$_{3}$, we have found
that $\Delta $H$_{pp}$(T) and $\chi _{esr}$(T) follow closely the behavior
of the monoclinic angle\cite{Alonso1} $\beta $(T) shown for comparison in
the inset of Fig. 4. The correlation of $\Delta $H$_{pp}$(T) and $\beta $(T)
is a consequence of the change of the site symmetry: an increase of the site
symmetry is usually accompanied by a narrowing of the ESR line.\cite{Tovar}

Our interpretation of the experiments is then consistent with the existence
of two different Ni sites in the monoclinic structure.\cite{Alonso1} If Ni1
(expanded) sites were occupied by Ni$^{2+}$ and Ni2 (contracted) sites by Ni$%
^{4+}$ one would expect Ni-O distances\cite{Juan2} of 2.06 \AA\ and 1.88 \AA
, respectively. Alonso et al.\cite{Alonso1} have determined that across the
structural transition at T$_{MI}$ the Ni-O distances pass from 1.958 \AA\ to
1.994 \AA\ and 1.923 \AA\ for the two different sites. Based on the size of
the octahedra they propose different effective valence states: +2.65 and
+3.35 for the Ni ions, representing the stabilization of an incomplete
charge disproportionation of Ni$^{3+}$ into Ni$^{2+}$ and Ni$^{4+}$. Their
observation points in the same direction of our results.

In summary, we have studied the magnetic properties of the Ni lattice in YNiO%
$_{3}$. In the metallic phase, at variance with LaNiO$_{3}$, we observed a
T-dependent susceptibility that we associate with the small Y$^{3+}$ ionic
radius. In the insulator phase, the $\chi _{dc}$ behavior allow us to
discard the possibility of Ni$^{3+}$ $e_{g}$-electron localization. In this
phase, we have observed that Ni ions become resonant and the characteristics
of the ESR spectrum correspond to Ni$^{2+}$. The integrated intensity
indicates that only a portion of the Ni ions are resonant. Our results are
then consistent with a charge disproportionation of the type 2Ni$%
^{3+}\rightarrow $ Ni$^{2+}$+Ni$^{4+}$, accompanying the MI transition. In
this description the Ni$^{4+}$ ions contribute to $\chi _{dc}$ but are not
detected by ESR techniques.

\section{Aknowledgement}

We acknowledge J. Rodr\'{\i}guez-Carvajal and B. Alascio for helpful
comments. This work was partially supported by ANPCyT (Argentina) through
the PICT 03-05266, and by the Ministry of Science and Technology (Spain)
under the project MAT2001-0539. R. D. S. is CONICET (Argentina) researcher.

\bigskip

{FIGURE CAPTIONS} \bigskip

{\noindent FIG. 1: }Temperatures T$_{MI}$ (open circles) and T$_{N}$ (open
triangles) vs. ionic radius r$_{R}$ for RNiO$_{3}$ taken from the literature
(Refs. 1, 5, and 20). With solid symbols we show the results for YNiO$_{3}$
(present work) and for the La$_{1-x}$Eu$_{x}$MnO$_{3}$ series (Ref. 13).
Solid lines are eyes guide. The dashed line separates regions where
T-dependent ( $\chi _{CW}$) and T-independent ($\chi _{0}$) susceptibilities
are observed, as explained in the text.

\smallskip

{FIG. 2: (a) Temperature dependence of }$\chi _{{dc}}${(T) (solid lines) and
}$\chi _{{esr}}${(T) (circles). The arrows indicate the AFM and MI
transition temperatures. Dashed line is an eyes guide. In the inset }$\chi _{%
{dc}}^{{-1}}${(T) is shown. }

\smallskip

FIG. 3: ESR spectra taken at different temperatures. Open circles for T =
560 K show the line obtained after subtracting spurious cavity spectra.

\smallskip
FIG. 4: Temperature dependence of the linewidth $\Delta $H$_{pp}$(T). In the
inset we show the monoclinic angle $\beta $(T) vs. T taken from Ref. 3.


\begin{references}
\bibitem{Nota}  The g-factor for Ni$^{2+}$ is (see Ref. Bleaney) g = 2.0023
- $\lambda $/$\Delta $. The value g = 2.16 is obtained using $\lambda $ =
-242 cm$^{-1}$ for the spin-orbit constant and $\Delta $ = 12000 cm$^{-1}$
for the energy gap to the first excited level (estimated from Ni$^{2+}$ in
diamagnetic oxides).
\end{references}
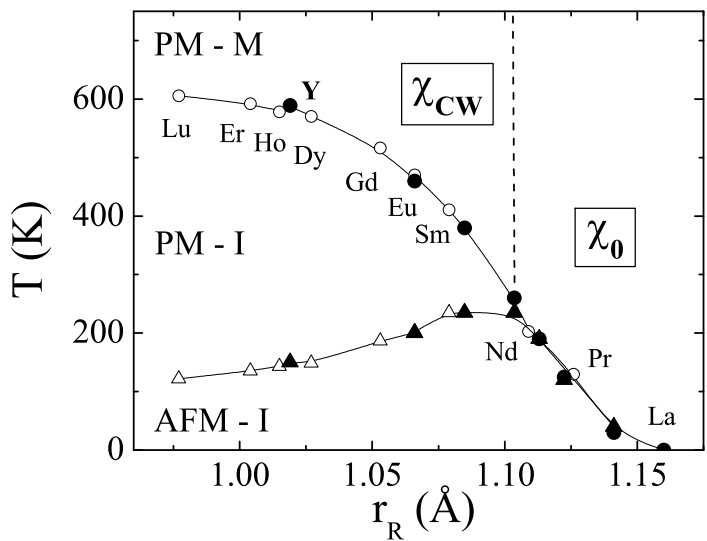
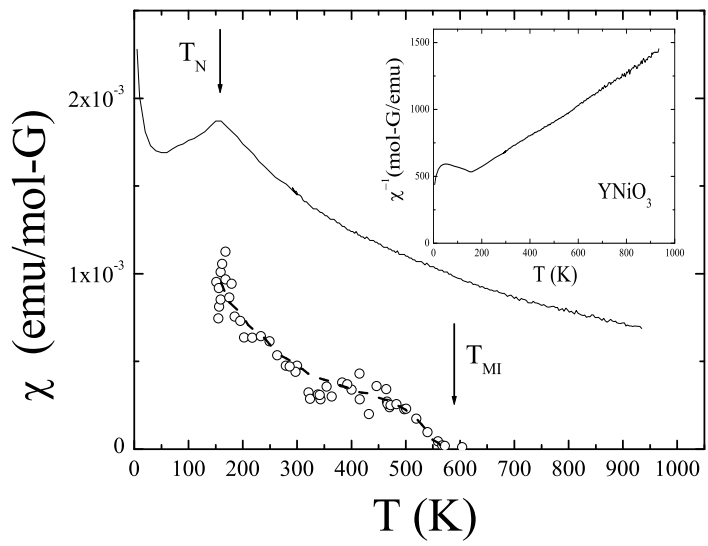
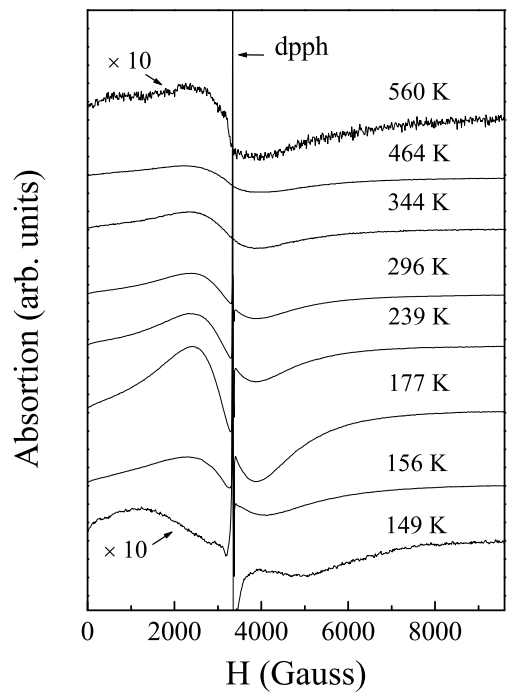
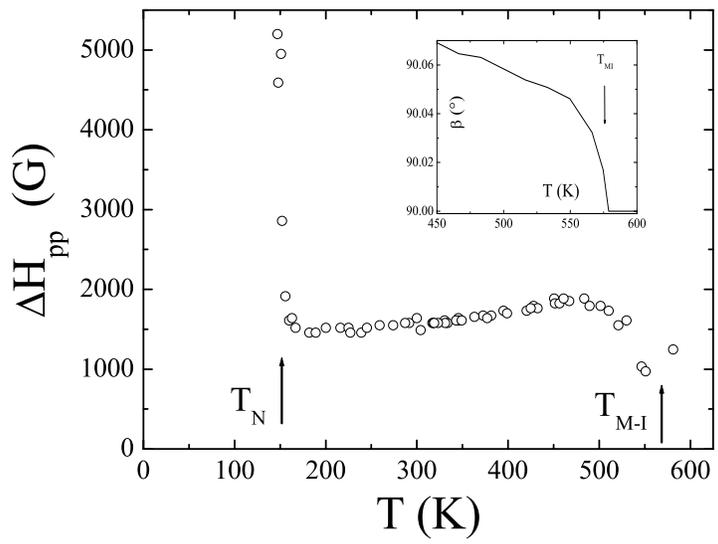
\end{document}